\documentclass[journal,twoside,web]{IEEEtran}
\usepackage{lcsys}
\usepackage{cite}
\usepackage{amsmath,amssymb,amsfonts}
\usepackage{algorithmic}
\usepackage{graphicx}
\usepackage{textcomp}

\usepackage[utf8]{inputenc}
\usepackage{amsmath}
\usepackage{amssymb}
\usepackage{graphicx,balance}
\usepackage{pifont}
\usepackage{bm}
\usepackage{dsfont}
\usepackage{subcaption}

\usepackage{siunitx}

\usepackage{amsfonts}
\usepackage{url}
\usepackage[pdftex,plainpages = false, colorlinks=true, linkcolor=black, citecolor = black, urlcolor = blue,pagebackref=false,hypertexnames=false, plainpages=false, pdfpagelabels]{hyperref}
\usepackage{balance}
\usepackage[capitalize]{cleveref}

\newtheorem{theorem}{Theorem}[section]

\crefname{section}{Section}{Sections}
\crefname{theorem}{Theorem}{Theorems}
\crefname{lemma}{Lemma}{Lemmas}
\crefname{table}{Table}{Tables}
\crefformat{equation}{(#2#1#3)}
\crefname{algocf}{Algorithm}{Algorithms}
\Crefname{algocf}{Algorithm}{Algorithms}
\crefname{ALC@unique}{Line}{Lines}

\newcommand{\x}{\mathbf{x}}

\newcommand{\you}{\mathbf{u}}
\newcommand{\picl}{{\pi}}
\newcommand{\fcl}{f_{cl}}
\newcommand{\rsoa}[1]{\bar{\mathcal{R}}_{#1}}
\newcommand{\srsoa}[1]{\bar{\mathcal{R}}_{#1}^s}
\newcommand{\Fcl}[1]{\bm{F}_{cl}^{#1}}
\newcommand{\crsoa}{\bar{\mathcal{R}}^c}
\newcommand{\lFcl}[1]{\underline{\bm{F}}_{cl}^{#1}}
\newcommand{\uFcl}[1]{\overline{\bm{F}}_{cl}^{#1}}
\newcommand{\safe}{\mathtt{s}}

\newcommand{\exper}[1]{$\mathsf{#1}$}
\newcommand{\experc}[2]{$\mathsf{#1}_{#2}$}
\newcommand{\meth}[1]{\textit{\texttt{#1}}}

\newcommand{\z}{\mathbf{z}}

\usepackage{accents}

\usepackage{paralist}

\crefformat{chapter}{\S#2#1#3}
\crefmultiformat{chapter}{\S\S#2#1#3}{and~#2#1#3}{, #2#1#3}{, and~#2#1#3}
\crefformat{section}{\S#2#1#3}
\crefmultiformat{section}{\S\S#2#1#3}{and~#2#1#3}{, #2#1#3}{, and~#2#1#3}

\usepackage{algorithm,algorithmic}

\def\BibTeX{{\rm B\kern-.05em{\sc i\kern-.025em b}\kern-.08em
    T\kern-.1667em\lower.7ex\hbox{E}\kern-.125emX}}
\begin{document}
\title{Constraint-Aware Refinement for Safety Verification of Neural Feedback Loops}
\author{Nicholas Rober, \IEEEmembership{Student Member, IEEE} and Jonathan P.\ How, \IEEEmembership{Fellow, IEEE}
\thanks{Submitted 09/12/2024. Research supported by Ford Motor Company.}
\thanks{N.\ Rober and J.\ How are with the Aerospace Controls Lab, Department of Aeronautics and Astronautics, Massachusetts Institute of Technology, Cambridge, MA 02319 USA (e-mail: nrober@mit.edu, jhow@mit.edu).}}

\maketitle

\begin{abstract}
Neural networks~(NNs) are becoming increasingly popular in the design of control pipelines for autonomous systems.
However, since the performance of NNs can degrade in the presence of out-of-distribution data or adversarial attacks, systems that have NNs in their control pipelines, i.e., neural feedback loops~(NFLs), need safety assurances before they can be applied in safety-critical situations.
Reachability analysis offers a solution to this problem by calculating reachable sets that bound the possible future states of an NFL and can be checked against dangerous regions of the state space to verify that the system does not violate safety constraints.
Since exact reachable sets are generally intractable to calculate, reachable set over approximations~(RSOAs) are typically used. The problem with RSOAs is that they can be overly conservative, making it difficult to verify the satisfaction of safety constraints, especially over long time horizons or for highly nonlinear NN control policies.
Refinement strategies such as partitioning or symbolic propagation are typically used to limit the conservativeness of RSOAs, but these approaches come with a high computational cost and often can only be used to verify safety for simple reachability problems. 
This paper presents Constraint-Aware Refinement for Verification (CARV): an efficient refinement strategy that reduces the conservativeness of RSOAs by explicitly using the safety constraints on the NFL to refine RSOAs only where necessary.
We demonstrate that CARV can verify the safety of an NFL where other approaches either fail or take up to 60x longer and 40x the memory.

\end{abstract}


\section{Introduction}
\label{sec:introduction}
\IEEEPARstart{N}{eural} networks~(NNs) have been applied to a wide range of control applications, including self-driving cars~\cite{chen2015deepdriving}, social navigation~\cite{zhu2021deep}, and control of drones in ground effect~\cite{shi2019neural}.
However, NNs can also perform poorly in the presence of adversarial attacks~\cite{yuan2019adversarial} or other out-of-distribution data~\cite{amodei2016concrete}.
When NNs are incorporated in the control loop of an autonomous system, i.e., a neural feedback loop~(NFL), errors from the NN can have undesireable compounding effects on the behavior of the system.
Thus, before NFLs can be applied in safety-critical situations where the health of people, the environment, and/or the system are at risk, safety assurances must be developed to detect such errors.

\begin{figure}[t]
    \centering
    \includegraphics[width=\columnwidth]{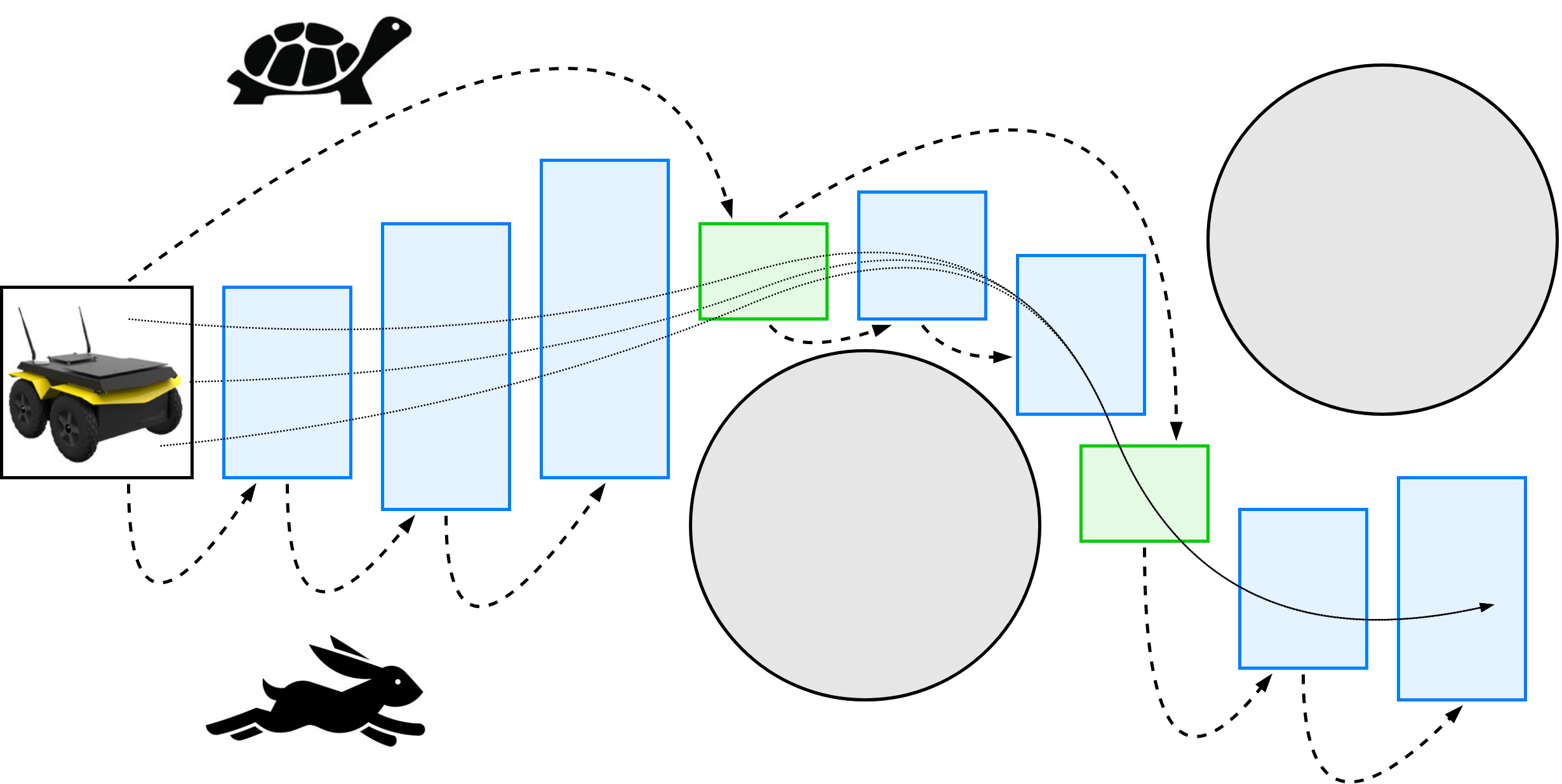}
    \caption{To verify that the robot avoids the obstacles (gray), CARV makes fast but conservative reachable set over-approximations (blue) unless they conflict with an obstacle, which prompts CARV to refine them with slower but less conservative approximations (green).}
    \vspace{-12pt}
    \label{fig:CARV_cartoon}
\end{figure}

Reachability analysis~\cite{huang2019reachnn,dutta2019reachability,everett2021reachability,chen2023one,sidrane2022overt,rober2023backward} addresses the problem of generating safety assurances for NFLs by calculating \emph{reachable sets} that provide bounds on the future states over a specified time horizon when the system starts in a given initial state set. 
The reachable sets can then be used to verify that the system does not violate the safety constraints of the system by checking for collisions between the reachable sets and the dangerous regions of the state space.
Since calculating exact reachable sets for NFLs is NP-hard~\cite{parrilo2001minimizing}, reachable set over-approximations~(RSOAs) offer a tractable alternative that can still be used to verify safety because they are supersets of the exact reachable sets.
While RSOAs are faster to calculate, they can be overly conservative, making it a challenge to verify safety because a conservative RSOA may indicate a constraint violation even when the system is in fact safe.
Thus, several approaches have been developed to \emph{refine} RSOAs to make them less conservative.

Partitioning~\cite{everett2021reachability,everett2020robustness,rober2023hybrid,xiang2018specification} accomplishes refinement by splitting up the initial state set and calculating reachable sets for each of the resulting subsets, which allows for tighter relaxations of the NN and thus less conservative RSOAs.
While partitioning is an effective approach for some problems, splitting up the initial set is a strategy that scales poorly with the state dimension of the NFL.

Another approach to refinement lies in symbolic reachability calculations~\cite{chen2023one}.
Symbolic RSOA calculations generate bounds on states $N > 1$ time steps in the future, thus mitigating the \emph{wrapping effect}~\cite{le2009reachability} where conservativeness is compounded by repeatedly taking over-approximations of over-approximations.
However, since an $N$-step calculation involves analyzing $N$ closed-loop time steps, symbolic calculations are very difficult for long time horizons.
Sidrane et al.~\cite{sidrane2022overt} overcame this challenge by alternating between symbolic and concrete calculations on a predefined schedule.
Recent work~\cite{sidrane2024ttt} builds on this strategy by determining a hybrid-symbolic schedule given a specified time budget.
While \cite{sidrane2022overt,sidrane2024ttt} tractably refine RSOAs over a given time horizon, they do not consider the safety constraints on the NFL and thus may not refine the RSOAs in a way that verifies safety.

To address the problem of tractable safety verification of NFLs, this paper presents the following contributions:
\begin{itemize}
    \item Constraint-Aware Refinement for Verification~(CARV): an algorithm that explicitly uses the system's constraints to guide the safety verification process for NFLs.
    \item A refinement algorithm that uses a hybrid-symbolic approach to avoid expensive RSOA calculations while still mitigating the wrapping effect to enable efficient safety verification for NFLs.
    \item Experiments wherein CARV verifies safety for a problem where other approaches either fail, are intractable, or take up to 60$\times$ longer and require 40$\times$ more memory.
\end{itemize}



\section{Preliminaries}
\label{sec:preliminaries}

\subsection{NFL System Dynamics}
\label{sec:nfl_dynamics}
The dynamics of a discrete-time system can be written as
\begin{equation}
    \label{eqn:general_dynamics}
    \x_{t+1} = f(\x_t, \you_t),
\end{equation}
where $\x_t \in \mathcal{X} \subseteq \mathbb{R}^n$ is the system's state, $\you_t \in \mathcal{U} \subseteq \mathbb{R}^m$ is the system's control input, and $f: \mathbb{R}^n \times \mathbb{R}^m \rightarrow \mathbb{R}^n$ is the system's discrete-time update function.
When the control input is designed as a function of the state, i.e., $\you_t = \picl(\x_t)$, the dynamics can be described in a closed-loop form as
\begin{equation}
    \label{eqn:nfl_dynamics}
    \x_{t+1} = \fcl(\x_t; \picl),
\end{equation}
where $\picl : \mathbb{R}^n \rightarrow \mathbb{R}^m$ is a NN and $\fcl: \mathbb{R}^n \rightarrow \mathbb{R}^n$.
Going forward, we will omit the implicit argument $\picl$ for brevity.
We consider the case where the system has a constraint function $c: \mathbb{R}^n \rightarrow [\mathtt{true},\ \mathtt{false}]$ that defines the set of safe states $ \mathcal{C} \triangleq \{\x \ \vert \ c(\x) = \mathtt{true} \} \subseteq \mathcal{X}$.

\subsection{Computational Graph Robustness Verification}
\label{sec:nn_verification}
Computational graphs~(CGs) are directed acyclic graphs that can be used to represent a series of computations, including NNs.
Given a computational graph $\bm{G}$ and input $\z \in \mathbb{R}^{n_i}$, we denote the output of the CG as $\bm{G}(\z) \in \mathbb{R}^{n_o}$.
The following theorem provides a result that will be used to calculate RSOAs.
\begin{theorem}
[CG Robustness~\cite{xu2020automatic}] \label{thm:lirpa}
Given a CG $\bm{G}$ and a hyper-rectangular set of possible inputs $\mathcal{I}$, there exist two explicit functions 
\begin{equation*}
    \underline{\bm{G}}(\z) = \bm{\Psi}\z + \bm{\alpha},\quad \overline{\bm{G}}(\z) =  \bm{\Phi}\z + \bm{\beta}
\end{equation*}
such that the inequality $\underline{\bm{G}}(\z) \leq \bm{G}(\z) \leq \overline{\bm{G}}(\z)$ holds element-wise for all $\z \in \mathcal{I}$,
with $\bm{\Psi}, \bm{\Phi} \in \mathbb{R}^{n_o \times n_i}$ and $\bm{\alpha}, \bm{\beta} \in \mathbb{R}^{n_o}$.
\end{theorem}

\subsection{Symbolic vs. Concrete Reachability}
\label{sec:reachability}
The reachable set of a system \cref{eqn:nfl_dynamics} at time $t$ given an initial state set $\mathcal{X}_0$ is defined as
\begin{equation}
    \mathcal{R}_{t}(\mathcal{X}_0) = \{\x \ \vert \ \x = \fcl^t(\x_0),\ \x_0 \in \mathcal{X}_0\},
\end{equation}
where $\fcl^t(\x_0) \triangleq \overbrace{\fcl \circ \ldots \circ \fcl}^{t}(\x_0)$ denotes $t$ compositions of $\fcl$.
Since exact reachability calculations for dynamics of the form \cref{eqn:nfl_dynamics} are generally intractable, we instead calculate RSOAs $\rsoa{t} \supseteq \mathcal{R}_{t}$ by using \cref{thm:lirpa} on a CG $\Fcl{k}$ (denoted $\Fcl{}$ when $k=1$) that takes input $\x_t$ and has output $\x_{t+k}$.

\begin{figure}[t]
    \centering
    \includegraphics[width=\columnwidth]{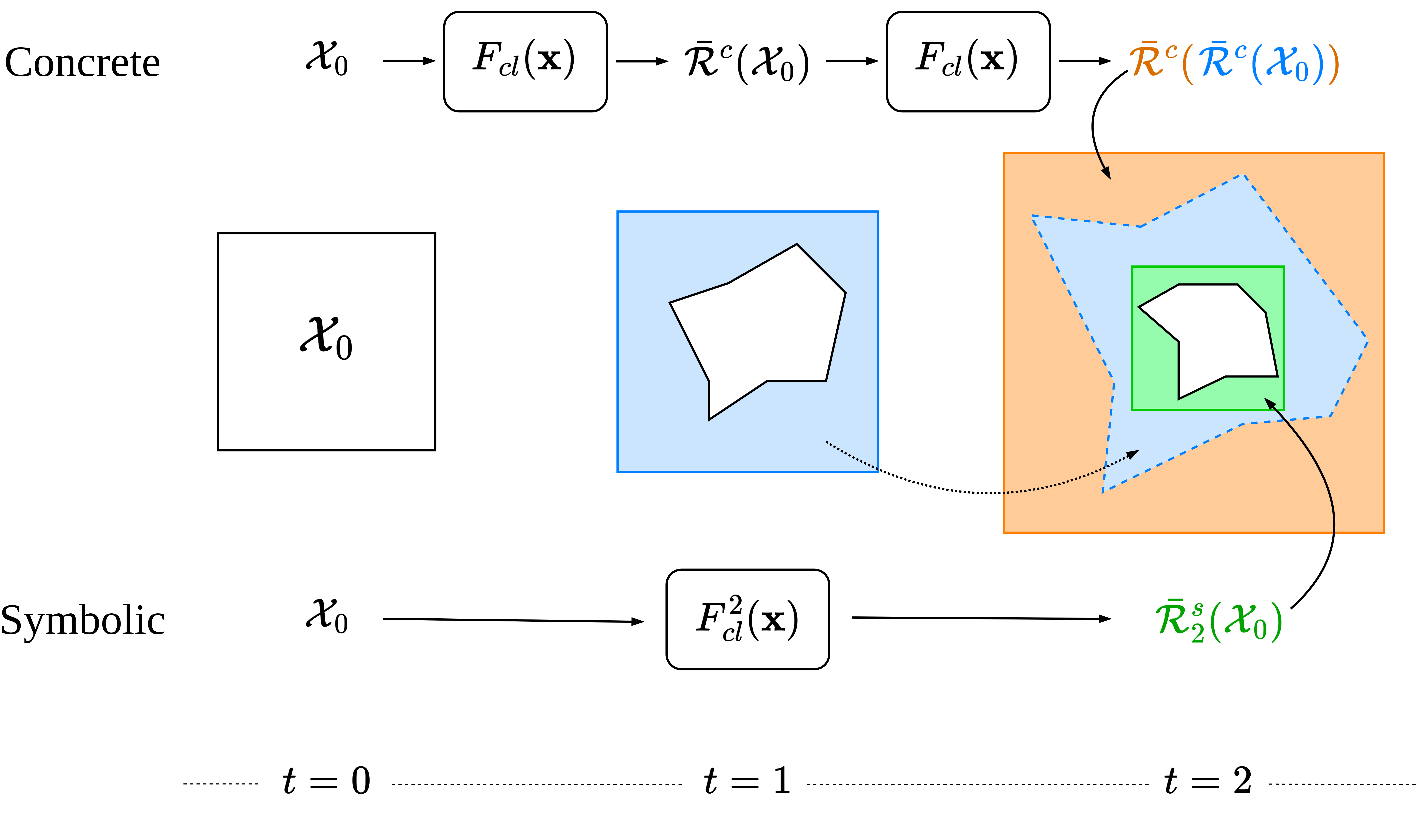}
    \caption{
    Concrete RSOA calculations (top) are subject to the wrapping effect~\cite{le2009reachability}.
    The concrete RSOA ${\crsoa(\crsoa(\mathcal{X}_0))}$ (orange) is an over-approximation of the true reachable set (blue, $t=2$) of the concrete RSOA at $t=1$.
    Symbolic RSOA calculations (bottom) use multiple self-compositions of $\fcl$ to calculate the RSOA at time $t=2$ (green) without having to calculate an RSOA at $t=1$, thus avoiding the wrapping effect.}
    \vspace{-14pt}
    \label{fig:concrete_vs_symbolic}
\end{figure}

In this paper we delineate between two types of reachability calculations: concrete and symbolic, shown in \cref{fig:concrete_vs_symbolic}.
Concrete (or one-step) RSOAs are defined as 
\begin{equation} 
    \label{eqn:concrete_rsoas}
    \crsoa(\rsoa{t}) = \{\x \ \vert \ \lFcl{}(\x_t) \leq \x \leq \uFcl{}(\x_t),\ \x_t \in \rsoa{t} \}
\end{equation}
where $\lFcl{}(\x_t)$ and $\uFcl{}(\x_t)$ are obtained from \cref{thm:lirpa} with CG $\Fcl{}$ and input set $\rsoa{t}$.
Notice that to approximate $\mathcal{R}_t(\mathcal{X}_0)$ with \cref{eqn:concrete_rsoas}, we need to calculate concrete RSOAs for each step prior to $t$, e.g, ${\mathcal{R}_2(\mathcal{X}_0) \subseteq \crsoa(\crsoa(\mathcal{X}_0))}$.
This approach introduces excess conservativeness due to the wrapping effect~\cite{le2009reachability}, which is shown in \cref{fig:concrete_vs_symbolic}.
The set $\crsoa(\crsoa(\mathcal{X}_0))$ (orange) is an over-approximation of the true reachable set (blue, $t=2$) of the RSOA from the prior time step (blue, $t=1$).
Since the RSOA at $t=1$ captures states that are not in the true reachable set at $t=1$, this conservativeness is passed onto the next time step.

Symbolic RSOA calculations can avoid the wrapping effect by approximating $\mathcal{R}_t(\mathcal{X}_0)$ directly.
We denote symbolic RSOAs with the notation
\begin{equation} 
    \label{eqn:symbolic_rsoas}
    \srsoa{t}(\mathcal{X}_0) = \{\x \ \vert \ \lFcl{t}(\x_0) \leq \x \leq \uFcl{t}(\x_0),\ \x_0 \in \mathcal{X}_0 \}.
\end{equation}
Since calculating $\lFcl{t}$ and $\uFcl{t}$ requires analyzing a CG that contains $t$ iterations of $\fcl$, symbolic RSOAs take much longer to generate as $t$ increases.

Thus, concrete and symbolic RSOA calculations each have their tradeoffs: concrete calculations are fast, but suffer from being overly conservative over multiple time steps whereas symbolic calculations are slower over long time horizons, but are much less conservative.
Given the challenges associated with these tradeoffs, the problem this paper addresses is how to efficiently verify that the state $\x$ of an NFL \cref{eqn:nfl_dynamics} stays in the safe region $\mathcal{C}$ of the state space over a given time horizon $t_f$.


\section{Constraint-Aware Refinement for Verification}
\label{sec:CARV}
This section presents Constraint-Aware Refinement for Verification (CARV).
The key insight behind CARV is that it does not matter if an RSOA is overly conservative as long as it does not conflict with the unsafe region $\mathcal{C}^\mathsf{c}$ (i.e., the complement of $\mathcal{C}$).
Thus, CARV's approach is to calculate concrete RSOAs until one violates the system's constraints, then refine the violating RSOA by recalculating it symbolically.
CARV addresses the problem of calculating symbolic RSOAs over long time horizons by setting a maximum symbolic horizon and prioritizing symbolic calculations with short horizons during the refinement step, as is described in more detail below.
Note that in this paper we only outline CARV for forward reachability, but it could be implemented for backward reachability~\cite{rober2023backward} using a similar approach.
\begin{figure}[t]
\vspace{-12pt}
\begin{algorithm}[H]
    \caption{CARV}
    \begin{algorithmic}[1]
        \setcounter{ALC@unique}{0}
        \renewcommand{\algorithmicrequire}{\textbf{Input:}}
        \renewcommand{\algorithmicensure}{\textbf{Output:}}
        \REQUIRE dynamics $\fcl(\cdot)$, constraints $c(\cdot)$, initial state set $\mathcal{X}_0$, time horizon $t_f$, maximum symbolic horizon $k_{max}$
        \ENSURE safety verification Boolean $\safe$
        \STATE $\safe \leftarrow \mathtt{true}$
        \STATE $\rsoa{0} \leftarrow \mathcal{X}_0$
        \FOR{$t$ in $\{1,\ \ldots,\  t_f\}$}
            \STATE $\rsoa{t} \leftarrow \mathtt{concrete\_reachability}(\rsoa{t-1},\ \fcl)$
            \IF[$\rsoa{t}$ violates constraint]{not $c(\rsoa{t})$} 
                \STATE $\rsoa{0:t} \leftarrow \mathtt{refine}(\rsoa{0:t},\ c,\ k_{max})$
            \ENDIF
            \IF[refined $\rsoa{t}$ violates constraint]{not $c(\rsoa{t})$}
                \STATE $\safe \leftarrow \mathtt{false}$
                \RETURN $\safe$
            \ENDIF
        \ENDFOR

        \RETURN $\safe$
    \end{algorithmic}\label{alg:CARV}
\end{algorithm}
\vspace{-20pt}
\end{figure}

\cref{alg:CARV} shows the pseudocode for CARV.
At each time step $t$ in the desired horizon, $\rsoa{t}$ is first calculated as a concrete RSOA using $\mathtt{concrete\_reachability}((\rsoa{t-1},\ \fcl)) \triangleq \crsoa(\rsoa{t-1})$.
If a collision is detected between $\rsoa{t}$ and the constraints from $c$, i.e., $c(\rsoa{t})$ is false, then CARV attempts to refine $\rsoa{t}$ until $c(\rsoa{t})$ is true.
If $\rsoa{t}$ still violates the constraints after refinement, the problem is not verified as safe and CARV will return $\mathtt{false}$.
Otherwise, if all RSOAs satisfy the constraints and CARV has reached the end of the time horizon, it will return $\mathtt{true}$, indicating that safety is verified.

\begin{figure}[t]
\vspace{-12pt}
\begin{algorithm}[H]
    \caption{refine}
    \begin{algorithmic}[1]
        \setcounter{ALC@unique}{0}
        \renewcommand{\algorithmicrequire}{\textbf{Input:}}
        \renewcommand{\algorithmicensure}{\textbf{Output:}}
        \REQUIRE reachable sets $\rsoa{0:t}$, constraints $c(\cdot)$, maximum symbolic horizon $k_{max}$
        \ENSURE reachable sets $\rsoa{0:t}$
        \STATE $t \leftarrow \rsoa{t}.t$
        \STATE $t_{min} \leftarrow \mathtt{max}(t - k_{max}, 0)$ \label{alg:refine:tmin}
        \STATE $k \leftarrow t - 2$
        \WHILE{$k \geq t_{min}$ and not $c(\rsoa{t})$}
            \IF{$\rsoa{k}.\mathtt{is\_symbolic}()$}
                \STATE $\rsoa{t} \leftarrow \mathtt{symbolic\_reachability}(\rsoa{k},\ \fcl,\ t - k)$
            \ELSIF{$k == t_{min}$}
                \STATE\COMMENT{refine $\rsoa{t}$ by taking $k_{max}$-sized steps to 0}
                \STATE $\rsoa{0:t} \leftarrow \mathtt{refine\_sequence}(\rsoa{0:t},\ k_{max})$
            \ENDIF
            \STATE $k \leftarrow k - 1$
        \ENDWHILE
        
        \RETURN $\rsoa{0:t}$
    \end{algorithmic}\label{alg:refine}
\end{algorithm}
\vspace{-12pt}
\end{figure}

The pseudocode for the $\mathtt{refine}$ algorithm is shown in \cref{alg:refine}.
The purpose of $\mathtt{refine}$ is to reduce the conservativeness of the most recently calculated RSOA $\rsoa{t}$ when it violates the safety constraints $c$.
Given the list of all previously calculated RSOAs $\rsoa{0:t} \triangleq \{\rsoa{0}, \ldots,\rsoa{t}\}$ and a maximum symbolic calculation horizon $k_{max} \in \mathbb{N}$, we loop backward until the RSOA at $t_{min}$ (defined \cref{alg:refine:tmin}) and try to deconflict $\rsoa{t}$ by refining it with symbolic RSOA calculations using $\mathtt{symbolic\_reachability}((\rsoa{k},\ \fcl,\ t-k)) \triangleq \srsoa{t-k}(\rsoa{k})$.
Thus, for each RSOA $\rsoa{k}$ in $\rsoa{t_{min}:t}$, if $\rsoa{k}$ is the result of a symbolic calculation (indicating $\rsoa{k}$ was previously refined), we try refining $\rsoa{t}$ with a symbolic calculation from $\rsoa{k}$.
If $\rsoa{t}$ still conflicts with $c$, we continue through the symbolic horizon until $t_{min}$.
If $t_{min}$ is reached and $\rsoa{t}$ still conflicts with $c$, (either because $\rsoa{t_{min}:t}$ are all the result of concrete calculations, or are simply not tight enough to deconflict $\rsoa{t}$), we call the function $\mathtt{refine\_sequence}$, shown in \cref{alg:refine_sequence}.

\begin{figure}
\vspace{-12pt}
\begin{algorithm}[H]
    \caption{refine\_sequence}
    \begin{algorithmic}[1]
        \setcounter{ALC@unique}{0}
        \renewcommand{\algorithmicrequire}{\textbf{Input:}}
        \renewcommand{\algorithmicensure}{\textbf{Output:}}
        \REQUIRE reachable sets $\rsoa{0:t}$, maximum symbolic horizon $k_{max}$
        \ENSURE reachable sets $\rsoa{0:t}$
        \STATE $t \leftarrow \rsoa{t}.t$
        \STATE $t_{min} \leftarrow \mathtt{max}(t - k_{max}, 0)$
        \IF{$t_{min} == 0$}
            \STATE $\rsoa{t} \leftarrow \mathtt{symbolic\_reachability}(\rsoa{0}, \fcl, t)$
        \ELSE
            \STATE $\rsoa{0:t_{min}} \leftarrow \mathtt{refine\_sequence}(\rsoa{0:t_{min}},\ k_{max})$
            \STATE\COMMENT{$\rsoa{t_{min}}$ is symbolic now}
            \STATE $\rsoa{t} \leftarrow \mathtt{symbolic\_reachability}(\rsoa{t_{min}}, \fcl, t - t_{min})$
        \ENDIF

        \RETURN $\rsoa{0:t}$
    \end{algorithmic}\label{alg:refine_sequence}
\end{algorithm}
\vspace{-20pt}
\end{figure}

The goal of $\mathtt{refine\_sequence}$ is to get an unconservative RSOA at $t$ given $k_{max}$.
By recursively taking $k_{max}$-sized steps backward through $\rsoa{0:t}$, $\mathtt{refine\_sequence}$ refines a series of symbolic RSOAs starting from zero and leading to $\rsoa{t}$.

\begin{figure*}[ht!]
    \centering
    \captionsetup[subfigure]{aboveskip=-2pt,belowskip=-2pt}
    \begin{subfigure}[t]{0.32\textwidth}
        \centering
        \captionsetup{width=.9\columnwidth}
        \includegraphics[width=1\columnwidth,trim={0 0 0 -1 cm },clip]{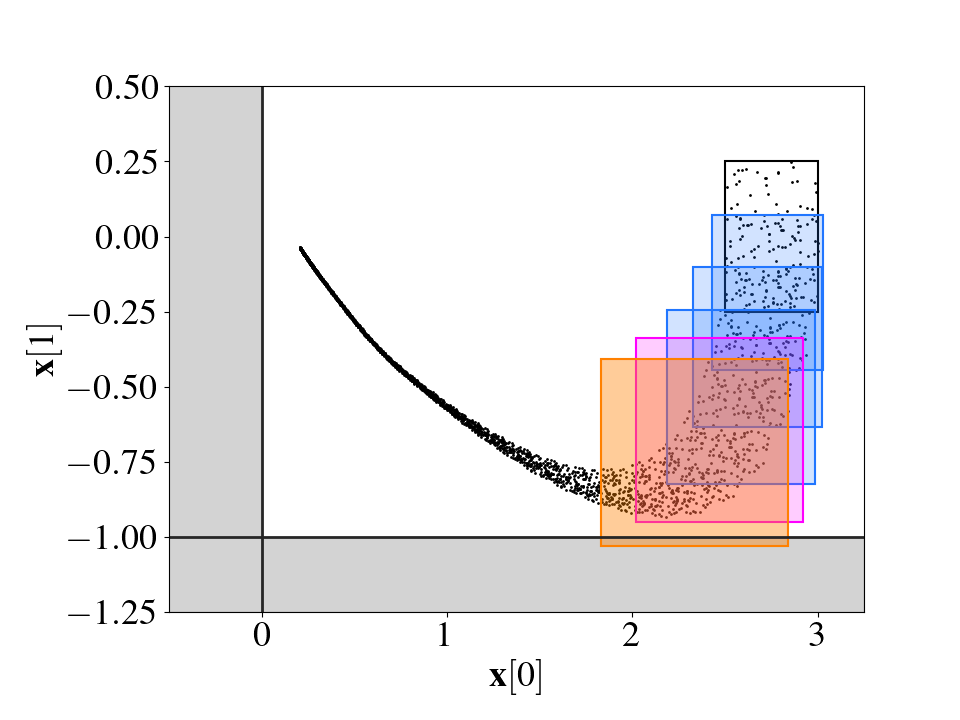}
        \caption{Concrete RSOAs (blue) are calculated until $\rsoa{5}$ (orange), which is calculated from $\rsoa{4}$ (magenta), violates the safety constraint (gray).}
        \label{fig:double_integrator:conflict}
    \end{subfigure}
    \begin{subfigure}[t]{0.32\textwidth}
        \centering
        \captionsetup{width=.9\columnwidth}
        \includegraphics[width=1\columnwidth,trim={0 0 0 0},clip]{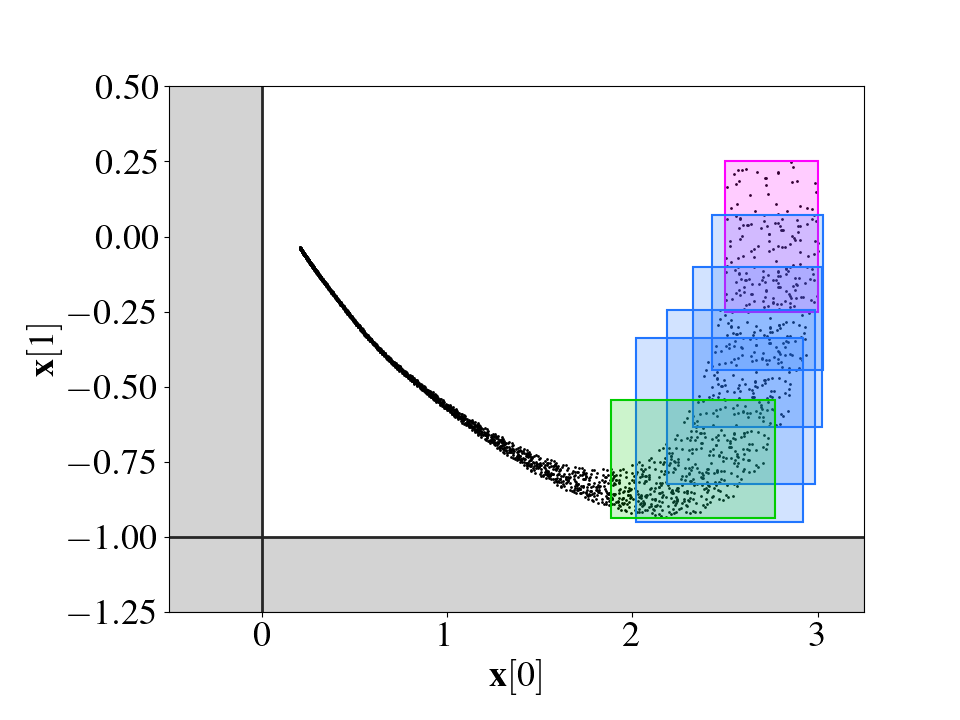}
        \caption{CARV uses $\mathtt{refine}$ (\cref{alg:refine}) to deconflict $\rsoa{5}$ with the constraint ${\x[0] \geq -1}$.
        The refined RSOA (green) is the result of a symbolic RSOA calculation from $\rsoa{0}$ (magenta).}
        \label{fig:double_integrator:resolved}
    \end{subfigure}
    \begin{subfigure}[t]{0.32\textwidth}
        \centering
        \captionsetup{width=.9\columnwidth}
        \includegraphics[width=1\columnwidth,trim={0 0 0 0},clip]{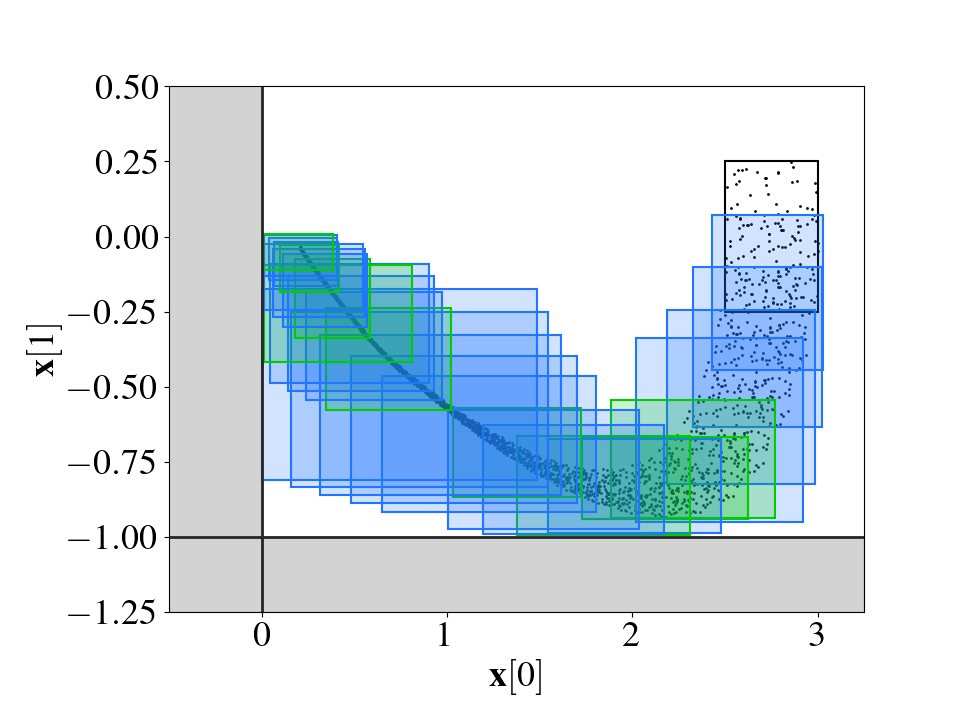}
        \caption{
        CARV calculates concrete RSOAs (blue) which can be very conservative unless they violate the contraints (gray), in which case the RSOA is refined with a symbolic calculation (green).
        }
        \label{fig:double_integrator:final}
    \end{subfigure}
    \caption{Successive frames of CARV's solution to \exper{DI} showing how CARV detects a collision (\cref{fig:double_integrator:conflict}), fixes it with refinement (\cref{fig:double_integrator:resolved}), and repeats until the problem is successfully verified (\cref{fig:double_integrator:final}).}
    \vspace{-12pt}
    \label{fig:double_integrator}
\end{figure*}

Note that while each individual RSOA calculation is symbolic, since $\mathtt{refine\_sequence}$ makes over-approximations of over-approximations like is done with concrete reachability, the wrapping effect is present, though reduced, in its result.
Larger values of $k_{max}$ allow $\mathtt{refine\_sequence}$  to take longer symbolic steps, thus reducing the magnitude of the wrapping effect in CARV.
The authors of \cite{sidrane2024ttt} propose a heuristic to determine $k_{max}$ based on a specified time budget, which is also a strategy that could be used for CARV.
However,  
the experimental results in \cref{sec:results} demonstrate that CARV's ability to verify a given problem is insensitive to the choice of $k_{max}$ and that, if CARV works for some minimum value $\underline{k}_{max}$, any $k_{max} \geq \underline{k}_{max}$ is also successful.
Thus, we leave the systematic determination of an appropriate $k_{max}$ to future work.

\section{Numerical Results}
\label{sec:results}
This section presents results from numerical experiments that demonstrate the properties of CARV.
We compare CARV to three existing refinement approaches: partitioning~\cite{everett2021reachability} (\meth{part}), pure symbolic propagation~\cite{chen2023one} (\meth{symb}), and hybrid-symbolic propagation~\cite{sidrane2022overt} (\meth{hybr}).
We show that CARV outperforms existing methods for a relatively simple double integrator example, as well as a more complicated ground robot example that requires safety verification over 52 time steps.
Reachable sets are generated with the AutoLiRPA CG analysis tool~\cite{xu2020automatic} using CROWN~\cite{zhang2018efficient} to generate bounds.
All experiments were conducted on a machine running Ubuntu 22.04 with an i7-6700K CPU and 32 GB of RAM.

\subsection{Double Integrator}
\label{sec:results:double_integrator}

First, we demonstrate CARV's behavior on a double integrator model
\begin{equation}
    \label{eqn:di_dynamics}
    \mathbf{x}_{t+1} =
    \begin{bmatrix}
    1 & dt \\
    0 & 1
    \end{bmatrix} \mathbf{x}_t +
    \begin{bmatrix}
    \frac{1}{2} dt^2 \\ dt
    \end{bmatrix} \mathbf{u}_t
\end{equation}
with $dt = 0.2$ and where $\you_t = \picl(\x_t)$ has 3 hidden layers with [30, 20, 10] neurons, ReLU activations, and was trained using an MPC expert to drive the state to ${\x = [0, 0]^\top}$ while satisfying the constraints
\begin{equation}
    \label{eqn:di_constraints}
    c(\x) = (\x[0] \geq 0\ \mathrm{and}\ \x[1] \geq -1),
\end{equation}
where $\x[0]$ and $\x[1]$ are the first and second states, respectively.
We denote the problem defined above as \exper{DI}.

\cref{fig:double_integrator} shows how CARV is able to verify safety against the constraints \cref{eqn:di_constraints} (gray) over ${t_f = 30}$ time steps with ${k_{max} = 15}$.
Monte Carlo samples (black markers) are propagated from the initial state set (black rectangle) as a proxy for the true behavior of the system.
In \cref{fig:double_integrator:conflict}, concrete RSOAs (blue) are calculated until $\rsoa{5}$ (orange) (calculated from $\rsoa{4}$, magenta) violates $c$.
\cref{fig:double_integrator:resolved} shows the result of calling $\mathtt{refine}$ on $\rsoa{5}$ in \cref{fig:double_integrator:conflict}: $\rsoa{5}$ (green) is refined as a symbolic RSOA from $\rsoa{0}$ (magenta).
\cref{fig:double_integrator:final} shows the result of iterating the processes shown in \cref{fig:double_integrator:conflict} and \cref{fig:double_integrator:resolved}: safety is successfully verified.
The example in \cref{fig:double_integrator} was calculated in \SI{3.71}{s}.

\subsection{Ground Robot}
\label{sec:results:ground_robot}

\begin{figure}[ht!]
    \centering
    \begin{subfigure}{0.97\columnwidth}
        \centering
        \includegraphics[width=1\columnwidth,trim={25 10 36 25pt},clip]{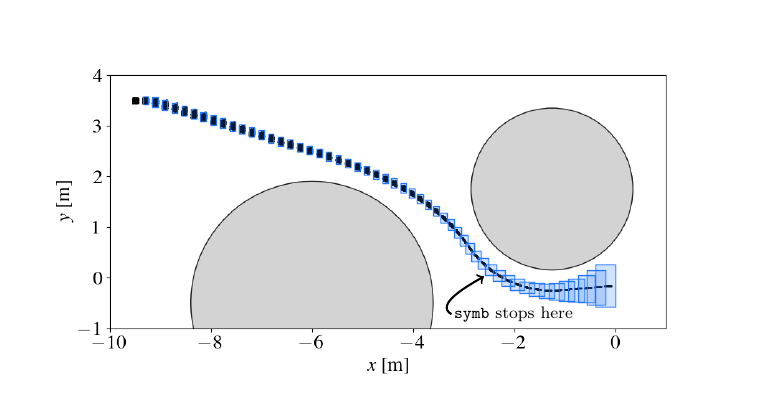}
        \caption{\meth{part} calculates partitioned concrete RSOAs (blue) to verify \exper{GR} as safe in \SI{540}{s} and uses 32 GB of RAM. 
        {\color{black}\meth{symb} calculates 40 symbolic RSOAs then runs out of memory at \SI{390}{s}.}}
        \label{fig:ground_robot:part}
    \end{subfigure}
    \begin{subfigure}{0.97\columnwidth}
        \centering
        \includegraphics[width=1\columnwidth,trim={25 10 36 25pt},clip]{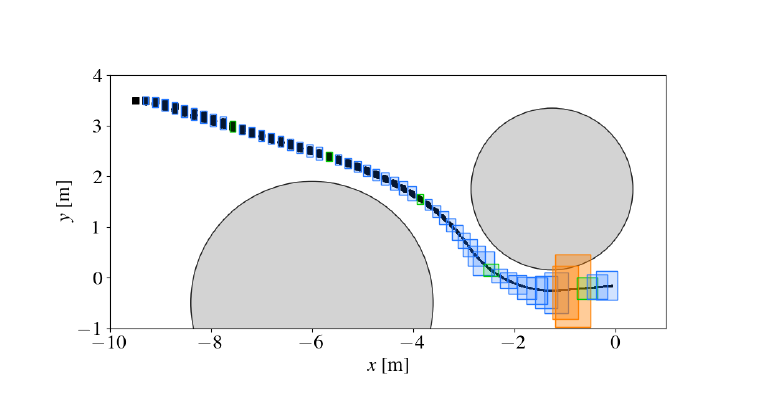}
        \caption{\meth{hybr} produces RSOAs (orange) that violate the safety constraints, and is thus unable to verify \exper{GR} as safe.}
        \label{fig:ground_robot:TTT}
    \end{subfigure}
    \begin{subfigure}{0.97\columnwidth}
        \centering
        \includegraphics[width=1\columnwidth,trim={25 10 36 25pt},clip]{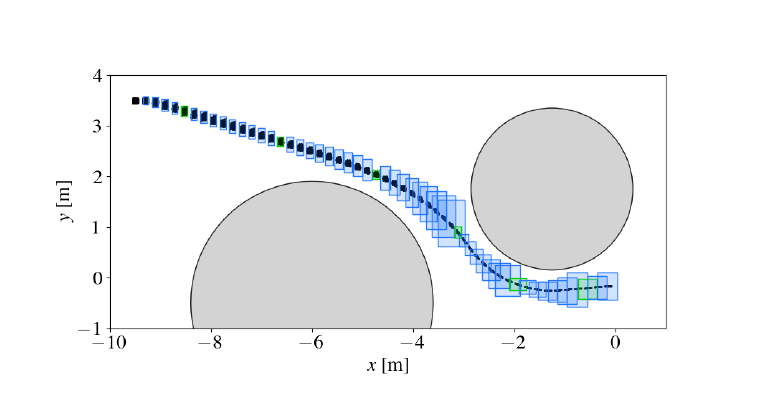}
        \caption{CARV verifies \exper{GR} as safe in \SI{13}{s} of computation time using a combination of concrete (blue) and refined (green) RSOAs.}
        \label{fig:ground_robot:CARV}
    \end{subfigure}
    \caption{CARV (ours) outperforms \meth{part}, \meth{symb}, and \meth{hybr} in computation time and ability to verify safety for \exper{GR}.}
    \label{fig:ground_robot}
    \vspace{-12pt}
\end{figure}

Next, we compare CARV to existing techniques with a verification problem for a ground robot with nonlinear dynamics
\begin{equation}
    \label{eqn:unicycle_dynamics}
    \mathbf{x}_{t+1} =
    \underbrace{
    \begin{bmatrix}
        x_t \\ y_t \\ \psi_t
    \end{bmatrix}}_{\x_t} +
    \begin{bmatrix}
        v\cos(\psi_t) \\ v\sin(\psi_t) \\ \omega_t
    \end{bmatrix} dt,
\end{equation}
where $(x_t,\ y_t)$ is the position of the robot, $\psi_t$ is it's heading, $v = 1$ is constant, $\omega_t$ is the input, and $dt = 0.2$.
The input $\omega_t$ is the output of a NN, i.e., $\omega_t = \picl(\x)$, where $\picl$ has 3 hidden layers with [40, 20, 10] neurons, ReLU activations, and was trained using an MPC expert to drive the position to ${(x, y) = (0, 0)}$ while avoiding two circular obstacles, i.e., satisfying the constraints
\begin{equation}
    \label{eqn:unicycle_constraints}
    \begin{split}c(\x) =  (
        & (x - c_{x1})^2 + (y - c_{y1})^2 \geq r_1^2 \ \mathrm{and} \\ & (x - c_{x2})^2 + (y - c_{y2})^2 \geq r_2^2)),
        \end{split}
\end{equation}
where $(c_{x1},\ c_{y1},\ r_1) = (-6,\ -0.5,\ 2.2)$ and $(c_{x2},\ c_{y2},\ r_2) = (-1.25,\ 1.75,\ 1.6)$ represent the $x-y$ positions and radii of the first and second obstacles, respectively.
We denote the problem above as \exper{GR} and show the corresponding outputs from \meth{part} (\cref{fig:ground_robot:part}), \meth{symb},
\meth{hybr} (\cref{fig:ground_robot:TTT}), and CARV (\cref{fig:ground_robot:CARV}) in \cref{fig:ground_robot}.

Partitioning (\meth{part}) is an approach where the initial state set is split into smaller subsets.
By doing RSOA calculations for each of the subsets, the CG relaxations $\lFcl{}$ and $\uFcl{}$ are taken over smaller regions and can thus be tighter.
As shown in \cref{fig:ground_robot:part}, this strategy can be used to verify \exper{GR} as safe.
The main problem with \meth{part} is that it scales poorly with the dimension of the state space.
In order to verify safety for \exper{GR} using \meth{part}, the initial state set had to be split into $6\times6\times18$ uniform partitions.
Since \exper{GR} needs to be verified over 52 time steps, this means \meth{part} makes 33696 concrete RSOA calculations, which takes \SI{540}{s} and uses all 32 GB of available RAM.
An additional challenge associated with \meth{part} is that an effective partitioning scheme is nontrivial to find.
While guided partitioning strategies exist \cite{everett2020robustness}, these approaches can take many iterations to converge, so we used trial and error to find a uniform partitioning configuration that worked.

Pure symbolic propagation (\meth{symb}) is another approach that can work well for small problems.
However, as noted in \cref{fig:ground_robot:part}, it {\color{black} scales poorly to} long time horizons.
In the case of \exper{GR}, the nonlinearites in the system and NN controller make for a challenging problem to calculate symbolically for many steps, causing {\color{black} our machine to run out of available memory (OOM Error) after \meth{symb} had calculated 40 RSOAs.
While the RSOAs produced by \meth{symb} were as tight as those from \meth{part}, \meth{symb} took \SI{390}{s} to calculate 40 RSOAs and is estimated to take \SI{864}{s} to verify \exper{GR} over the entire time horizon.}

Hybrid-symbolic propagation (\meth{hybr}) was introduced in \cite{sidrane2022overt} and recently employed as the refinement strategy for \cite{sidrane2024ttt}.
In hybrid symbolic propagation, symbolic RSOAs are calculated from previous symbolic RSOAs on a given schedule, with concrete RSOAs filling in the gaps.
While significant effort could be expended to find a hybrid schedule that works for any specific problem, we will consider a uniform schedule that makes symbolic calculations every $k_{max} = 10$ steps.
\cref{fig:ground_robot:TTT} shows the result produced by \meth{hybr} for \exper{GR}, which was calculated in \SI{9.71}{s}.
While much faster than \meth{part} and \meth{symb}, two RSOAs toward the end (orange) intersect with the obstacle, meaning that \meth{hybr} cannot verify \exper{GR} as safe in this configuration.
In \cref{sec:results:k_max} we investigate the importance of $k_{max}$ on this outcome.

Finally, the result of using CARV with $k_{max} = 10$ on \exper{GR} is shown in \cref{fig:ground_robot:CARV}.
Since none of the RSOAs in the final result intersect with the obstacle, CARV verifies that \exper{GR} is safe.
Moreover, it takes \SI{9.32}{s} to run, thereby also outperforming existing methods in calculation time.
Notice that similar to \meth{hybr}, CARV calculates symbolic RSOAs at $k_{max}$ intervals.
However, CARV only does this \emph{as needed} to save on time, and in fact, for \exper{GR}, calculates only concrete RSOAs up to $t=35$ before there is a conflict with the second obstacle, triggering a call to $\mathtt{refine\_sequence}$ that refines RSOAs at $k_{max}$ intervals tracing back to $\mathcal{X}_0$.
The results from each refinement method for \exper{GR} and \exper{DI} are summarized in \cref{tab:alg_comparison}.

\begin{table}[t]
    \centering
    \caption{CARV efficiently verifies safety for \exper{DI} and \exper{GR}. $^{\bm{*}}$indicates verification unsuccessful.}

    \normalsize
    \begin{tabular}{lcc}
        \hline
        Approach & \exper{DI} [s] & \exper{GR} [s] \\
        \hline
        \meth{part} \cite{everett2021reachability} & $20.79 \pm 1.23$ & $540.10 \pm 1.73$ \\
        \meth{symb} \cite{chen2023one} & $29.42 \pm 1.06$ & {\color{black}OOM Error} \\
        \meth{hybr} \cite{sidrane2022overt} & $1.77^{\bm{*}} \pm 0.14$ & $9.71^{\bm{*}} \pm 0.12$ \\
        CARV (\textbf{ours}) & $\mathbf{3.11 \pm 0.12}$ & $\mathbf{9.32 \pm 0.25}$ \\
        \hline
    \end{tabular}
    \vspace{-16pt}
    \label{tab:alg_comparison}
\end{table}

\subsection{Varying the Maximum Symbolic Horizon}
\label{sec:results:k_max}
\begin{figure}[t]
    \centering
    \includegraphics[width=0.82\columnwidth,trim={20pt 24pt 42pt 36pt},clip]{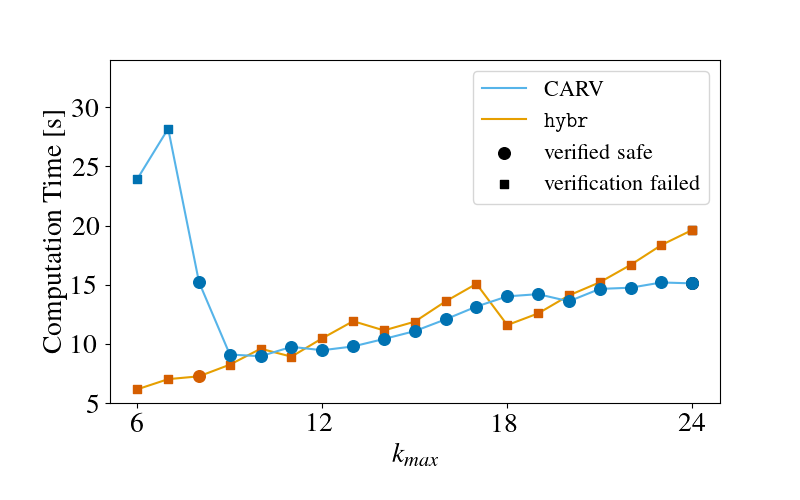}
    \caption{CARV verifies \exper{GR} as safe for all tested $k_{max} \geq 8$.}
    \label{fig:ground_robot_comparison}
    \centering
    \hspace*{12pt}\includegraphics[width=0.82\columnwidth,trim={20pt 24pt 42pt 36pt},clip]{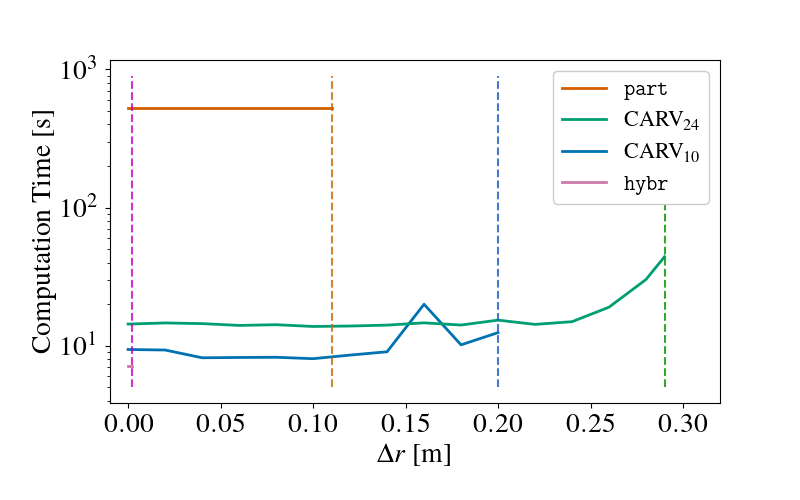}
    \vspace{-3pt}
    \caption{CARV can verify harder problems than \meth{hybr} or \meth{part}. Dashed vertical lines mark largest values of $\Delta r$ that can be verified by each algorithm.}
    \vspace{-12pt}
    \label{fig:sweep_constraint}
\end{figure}

Next we investigate CARV's sensitivity to the hyperparameter $k_{max}$ when applied to \exper{GR}.
\cref{fig:ground_robot_comparison} shows results from both \meth{hybr} and CARV given $k_{max}$ values ranging from 6 to 24.
For $k_{max} < 8$, symbolic calculations are short and frequent so neither \meth{hybr} or CARV can sufficiently reduce the wrapping effect, thus leading to a failure to verify \exper{GR} as safe.
At $k_{max} = 8$, both \meth{hybr} and CARV verify \exper{GR} as safe, but for all tested $k_{max} > 8$, \meth{hybr} fails to verify \exper{GR} as safe whereas CARV verifies safety in each case.
Notice CARV spends more time in the cases where verification fails ($k_{max} = 6,\ 7$) because it tries to refine as much as possible in an attempt to verify safety.
{\color{black}For $k_{max} > 8$, CARV completes the verification problem in a similar amount of time as \meth{hybr}, but is more successful because it selectively refines its solution with the explicit goal of safety verification.}



\subsection{Verification Analysis}
Finally, we investigate CARV's ability to verify more challenging problems by introducing the parameter $\Delta r$ and solving a series of problems \experc{GR}{\Delta r} with the same constraint function \cref{eqn:unicycle_constraints} as \exper{GR}, but with radii $r_1 + \Delta r$ and $r_2 + \Delta r$.
{\color{black}As the radii of the two obstacles increase, the conservativeness of an approach becomes more impactful in that it may cause the approach to fail to verify safety.}
\cref{fig:sweep_constraint} shows successful verifications from CARV, \meth{part}, and \meth{hybr} on different \experc{GR}{\Delta r} where $\Delta r \in [0, 0.29]$.
The dashed vertical lines mark the largest values of $\Delta r$ that can be verified by each algorithm.
We set $k_{max} = 8$ for \meth{hybr} (magenta) since that was the only value that verified \exper{GR}  (\cref{fig:ground_robot_comparison}), but \meth{hybr} fails with $\Delta r = 0.002$.
Since \meth{part} (orange) uses all available RAM to verify \exper{GR}, we use the same partitioning scheme as in \cref{sec:results:ground_robot}, which verifies each \experc{GR}{\Delta r} with $\Delta r \leq 0.11$.
CARV is tested with $k_{max}$ values 10 and 24, denoted $\mathrm{CARV}_{10}$ (blue) and $\mathrm{CARV}_{24}$ (green), respectively.
Unlike \meth{part} and \meth{hybr}, CARV adjusts to the difficulty of the problem and thus has lower calculation times for lower values of $\Delta r$.
Additionally, CARV verifies more difficult problems than \meth{part} or \meth{hybr} -- we achieve verification up to \experc{GR}{0.20} and \experc{GR}{0.29} with $\mathrm{CARV}_{10}$ and $\mathrm{CARV}_{24}$, respectively.
Note that for hyper-rectangular RSOAs, $\Delta r = 0.33$ is the highest verifiable value.

\section{Conclusion}
\label{sec:conclusion}
In this paper we presented CARV: a refinement strategy that allows for efficient safety verification of NFLs subject to constraints that define safe regions of the state space.
Specifically, CARV selectively refines RSOAs in an attempt to deconflict them with the given constraints on the system.
The key idea behind CARV is that it is acceptable for RSOAs to be overly conservative as long as they do not lead to a failure in the safety verification.
Thus, CARV calculates concrete (fast but conservative) RSOAs by default, then refines them with symbolic (slow but tight) RSOA calculations if they violate the safety constraints.
We demonstrated CARV on two verification problems, including a ground robot example with a verification horizon of 52 time steps, and compared CARV to several other refinement techniques.
CARV demonstrated faster computation time than the other approaches and was the only approach that tractably verified both problems as safe.

In future work, we plan to further investigate the effect of the maximum symbolic calculation horizon $k_{max}$ and find a systematic way to determine it.
Additionally, we will seek to prove that if CARV works for some minimum value $\underline{k}_{max}$, any $k_{max} \geq \underline{k}_{max}$ is also successful, as is seen in our experiments.

\bibliographystyle{IEEEtran}
\bibliography{refs}

\end{document}